\documentclass[%
 aip,
 amsmath,amssymb,floatfix,
 reprint,%
]{revtex4-1}

\usepackage{graphicx}% Include figure files
\usepackage{dcolumn}% Align table columns on decimal point
\usepackage{bm}% bold math
\usepackage[utf8]{inputenc}
\usepackage[T1]{fontenc}
\usepackage{mathptmx}
\usepackage{color}
\usepackage{multirow}
\newcommand{\etal}{\textit{et al}.}

\usepackage[hhmmss]{datetime}
\begin{document}

\preprint{AIP/123-QED}

%\preprint{}

\title{On structural and dynamical factors determining the integrated basin instability of power-grid nodes} %Title of paper




\author{Heetae Kim}
\email[]{hkim@utalca.cl}
 \affiliation{Department of Industrial Engineering, Universidad de Talca, Curic{\'o} 3341717, Chile}%
\affiliation{Asia Pacific Center for Theoretical Physics, Pohang 37673, Korea}%

\author{Mi Jin Lee}
\email[]{mj.mijin.lee@gmail.com}

\affiliation{Department of Physics, Inha University, Incheon 22212, Korea}
\affiliation{Department of Physics, Sungkyunkwan University, Suwon 16419, Korea}

\author{Sang Hoon Lee}
\email[Corresponding author: ]{lshlj82@gntech.ac.kr}
\affiliation{Department of Liberal Arts, Gyeongnam National University of Science and Technology, Jinju 52725, Korea}%

\author{Seung-Woo Son}
\email[Corresponding author: ]{sonswoo@hanyang.ac.kr}
\affiliation{Asia Pacific Center for Theoretical Physics, Pohang 37673, Korea}
\affiliation{Department of Applied Physics, Hanyang University, Ansan 15588, Korea}%

\date{\today}

\begin{abstract}
In electric power systems delivering alternating current, it is essential to maintain its synchrony of the phase with the rated frequency. The synchronization stability that quantifies how well the power-grid system recovers its synchrony against perturbation depends on various factors. As an intrinsic factor that we can design and control, the transmission capacity of the power grid affects the synchronization stability. Therefore, the transition pattern of the synchronization stability with the different levels of transmission capacity against external perturbation provides the stereoscopic perspective to understand the synchronization behavior of power grids. In this study, we extensively investigate the factors affecting the synchronization stability transition by using the concept of basin stability as a function of the transmission capacity. For a systematic approach, we introduce the integrated basin instability, which literally adds up the instability values as the transmission capacity increases. We first take simple 5-node motifs as a case study of building blocks of power grids, and a more realistic IEEE 24-bus model to highlight the complexity of decisive factors. We find that both structural properties such as gate keepers in network topology and dynamical properties such as large power input/output at nodes cause synchronization instability. The results suggest that evenly distributed power generation and avoidance of bottlenecks can improve the overall synchronization stability of power-grid systems.
\end{abstract}


\maketitle %\maketitle must follow title, authors, abstract and \pacs


\textbf{In modern society, power grids play an essential role as one of the most important infrastructures by providing electrical energy. As the structure and the operation strategy of power grids are becoming more complex, designing and managing the power grids for stable electricity supply are also becoming a harder challenge. The problem of course belongs to the field of electrical engineering with all of the complicated practical matters, but there have been significant endeavors to analyze it with only the most essential ingredients, starting from arguably the simplest one: the topology of power grids as mathematical objects represented by graphs or networks. Initiated solely from this structural aspect, the past decades have witnessed the drastic advancement in the field of power-grid research, powered by the theoretical and practical tools of network science combined with nonlinear dynamics. One of the most quintessential approaches is the stability analysis of synchronization among power-grid nodes. It is known that the transmission capacity affects the synchronization stability. In the light of the fluctuating transmission capacity in real power grids, we investigate what structural and dynamical factors make the dynamic stability of power grids resilient over a range of transmission capacity.}

\section{\label{sec:introduction}Introduction}

The synchronization stability of individual nodes in a power grid quantifies how much each node can sustain its synchrony against external disturbance to the node's dynamic variables~\cite{Menck:2013bk}. 
Based on the nodal synchronization stability, studies have revealed various topological and functional properties of power grids to explain what makes power grids solidly synchronized~\cite{Schultz2014detours, Menck:2014fn, Ji:2014ina, Belykh:2016dq, Nitzbon:2017fo, Ji:2018fz, Kim:2015kg,Kim:2016kd,Kim:2018do}. Those power-grid studies on synchronization stability utilize the capacity of transmission lines as a key parameter. The transmission capacity affects the synchronization stability of nodes, sometimes non-monotonically~\cite{Kim:2018do}. However, it has yet to be understood crystal clear how the unique transition patterns of the nodal synchronization stability are related to other network properties. 

In this study, we investigate the transition pattern of the nodes in two power-grid models: the 5-node motifs as theoretical basic building blocks of power grids and the IEEE 24-bus model~\cite{Subcommittee:1979ih,Grigg:1999he} as a representative practical case from the engineering point of view. We introduce an integrated measure of basin (in)stability to comprehensively capture the transition pattern of the synchronization stability and reveal that the large amount of power input/output and the topological gate-keeper structure can be general signatures of unstable nodes.

As a contribution to this Focus Issue, we begin our paper by briefly guiding the readers through the historical development of power-grid studies that utilize the network perspective over the past decades in Sec.~\ref{sec:overview}, on which our contribution in this work introduced at the end of the section is based. We hope that this mini-review helps to better understand the context of our work, compared to the introductory part of an ordinary journal paper. The main part of our contribution introduces the basic theoretical framework in Sec.~\ref{sec:methods} and the power-grid models in Sec.~\ref{subsec:powergridmodels}. We present the main results and implications in Sec.~\ref{sec:results}, and conclude our work in Sec.~\ref{sec:conclusion}. Appendix~\ref{sec:Appendix} details the parameters of the power grid used in this study.

\section{\label{sec:overview}Overview of network-based approaches}

\subsection{\label{sec:structure}Purely structural approaches}

In the early 2000s, the nascent field of network science~\cite{Strogatz:2001fp,Newman:2003da}, equipped with an explosively increasing amount of various types of real data, started to analyze an intrinsically networked system of power grids as well. At that time, researchers mainly focused on the structural properties, for instance, the resilience of power-grid networks under random failures or intentional attacks~\cite{Albert2000}. By taking this approach, Albert \textit{et al}.~\cite{PhysRevE.69.025103} estimated the vulnerability of the North American power grid in response to intentional attacks by removing nodes based on the number of connected neighbors (degree) and the amount of net power input of each node, and then compared the result to that from random attack. Similar approaches were conducted based on betweenness---the fraction of shortest paths between all pairs of nodes that go through the node of interest---to the power grids of the western United States~\cite{PhysRevE.69.045104}, Italy~\cite{Crucitti:2004gf}, the North America~\cite{Kinney:2005jy}, and the Western China~\cite{Xu:2009tz}. Crucitti \textit{et al}. used the link betweenness (the same concept as the aforementioned betweenness but assigned to transmission lines instead of nodes) to identify the critical lines and suggested to install a few new lines at specific locations to improve the robustness of Spanish, French, and Italian power grids~\cite{Crucitti:2005vd}.

These seminal studies successfully projected power grids on the framework of network science by applying elementary topological properties such as degree, betweenness, and size of connected components. However, as any other types of real-world network data went through, researchers began to realize the limitation of these highly simplified approaches. Take the betweenness, for instance. In an ideal world, of course, the shortest path assumed to be used in calculating the betweenness is the most efficient and desirable one for the transmission of electricity. In reality, though, it is known that the power produced by local generators is more likely to be transmitted to the nearest substations due to various practical reasons (in this case, the geographical constraint) that the simple network analysis cannot capture~\cite{Kinney:2005jy}. Therefore, it is absolutely required to bring such specific contexts of power grids before blindly applying only the topological part of network analysis.

\subsection{Direct current approximation to find critical lines}
\label{sec:DC}

A power grid is a special type of distribution network where electrical currents flow from power sources, i.e., power generators to power sinks, i.e., consumers. The direct current (DC) approximation assumes the DC flowing from the highest electric potential at the generators to the lowest electric potential at the consumers, just as water flows from the highest mountain top to the lowest valley. This assumption is of course not technically correct, as most infrastructural power grids in the world use the alternating current (AC) as a result of the famous current war between Nikola Tesla and Thomas Edison. However, the DC approximation has the merit in its simplicity without taking all of the complicated factors of phases and the rated frequency, so it has been widely used as the proxy of real-world power-grid failure due to overloading.

Simply put, when the amount of the current flow exceeds the capacity of a transmission line, the line is overloaded and thus fails. Estimating the current flow with the DC approximation, we can identify such ``critical'' lines and they potentially trigger the cascading failure~\cite{Arianos:2009ua,Bompard:2009ga,Dong:2010fx,Chen:2010fo,Labavic:2014fd,Schafer:2017cc}. The size of the resultant power system blackout is argued to follow the power-law distribution and it is reproduced by a model study~\cite{Dobson:2007fq}. Once a line is overloaded, the line can be accidentally or intentionally (by power grid operators) disconnected from the grid. In that case, the load should be redistributed to other lines.  
The pattern of the load redistribution is estimated by the line outage distribution factor~\cite{JiachunGuo:hx,Ronellenfitsch:2017jc} related to network properties.
Various more realistic aspects were adopted such as the power loss due to the heat dissipation~\cite{Yang:2017io}, the noisy power input~\cite{Nesti:2018he}, and the islanding phenomenon~\cite{Mureddu:2016dw}.

The cascading propagation in the aforementioned studies is represented as the series of discrete stationary states of distinct networks. For example, one can get a stationary state of an original intact network to identify the most critical line. Assuming that critical lines are broken due to the overload, it is possible to simulate the rebound effect from the modified network with the critical lines removed. By repeating this process until there is no more overloaded line, one can estimate the total size of cascading failure or power outage.  

\subsection{Dynamic behaviors with the AC approximation}
\label{sec:AC}

As mentioned in Sec.~\ref{sec:DC}, more detailed dynamic behaviors of power grids such as synchronization should be investigated with the full consideration of the current's synchronized phase oscillation, i.e., the AC approximation.
For instance, the current flow of each link at a synchronized steady state can be utilized to find the critical links~\cite{Lozano:2012gm,Witthaut2012,Rohden:2016jq,Rohden:2017bu,Li:2017fn,Manik:2017ex}.
The transient pattern of nodes' phase and frequency is also a crucial indicator of critical behaviors~\cite{Schafer:2018to}.

If finding the critical lines to prevent the power outage is the main purpose of the DC approximation, it is possible to set more sophisticated strategies to improve power grids' performance with this more realistic AC approximation. Rohden \etal~\cite{Rohden:2017bu} compared the strategies to heal the damaged power grid and Tchawou \etal~\cite{TCHAWOU:gr} suggested a control strategy to maintain the synchronization.
Li \etal~\cite{Li:2017fn} and Tchuisseu \etal~\cite{Tchuisseu:2017cu} tried to improve power grids' stability by optimization and dynamic demand control, respectively. 
In addition, Witthaut \etal~\cite{Witthaut:2015km} revealed that the cascading effect is not limited to nearby regions and the long-range effects can occur due to cascading.

Moreover, modern power systems are becoming more complex by the implementation of distributed renewable power sources as a part of smart-grid systems. In this context, researchers started to investigate the effect of the spatial or geographical distribution of power sources. For instance, studies revealed that decentralized power systems are more stable compared than centralized ones~\cite{Rohden:2012gl,Lee:2017dz,Odor:2018kf,Wolff:2018bd,Schafer:2018gn}. The control strategy to maintain the grid's synchronization was also investigated in decentralized systems~\cite{Schafer:2015vu}.
Some studies investigated the nontrivial response of power grids, so called Braess's paradox, which refers to the phenomenon that removing a connection can counterintuitively increase the overall transportation efficiency~\cite{Witthaut2012,TCHAWOU:gr}.

In the midst of those various topics, one would need measures to quantify the synchronization stability. In Sec.~\ref{subsec:basinstability}, we introduce one of such measures that captures researchers' interest, which has eventually become the cornerstone of our analysis in this work.

\subsection{\label{subsec:basinstability}The synchronization stability}

The AC flowing throughout power grids is supposed to be synchronized at the rated frequency (mostly 50 or 60 Hz, depending on the regional standard). In real power systems, however, the phase of the AC may experience unexpected disturbances that could hamper the systems' synchrony. It is challenging to analyze the synchronization stability due to its nonlinear nature. For theoretical model studies to proceed, as a standard procedure, we consider each node in power grids as a phase oscillator with the phase of the AC flowing through the node. 

One can perturb a node's phase and frequency in a power grid and then count how many times the power-grid system recovers its overall synchronous state. Menck \etal~\cite{Menck:2013bk} suggested this approach by measuring the probability of the synchronization recovery against nodal perturbations, which is called the \emph{basin stability}. Assuming the principle of equal a priori probabilities, the basin stability of a node also corresponds to the fraction of configuration space composed of the node's phase and frequency disturbance from which the system recovers its synchrony. Since the perturbation is applied to each node at a time, it is sometimes called the single-node basin stability for technicality~\cite{Schultz2014detours}, while the multiple-node basin stability~\cite{Mitra:2017jp} refers to the result of simultaneous perturbations to several nodes.

In terms of the basin stability, it is known that the specific locations such as the terminal of the network branch, so-called the dead-end or the dead-tree~\cite{Menck:2014fn}, or the detour path~\cite{Schultz2014detours} tend to cause lower synchronization stability than others. 
The bistability of power-grid nodes is also analyzed over the state space of nodes~\cite{Ji:2014ina,Belykh:2016dq}.
Based on the transient pattern in the state space, different categories of the power-grid nodes in terms of topological roles are analyzed to show their distinct diagnostics in the basin stability~\cite{Nitzbon:2017fo}.
From a practical point of view, numerical studies on basin stability require heavy computational resources due to its time complexity in exploring the state space, so modified versions were developed to estimate the basin stability~\cite{Ji:2018fz} and sometimes with setting the finite-time limit~\cite{Hellmann:2016js,Schultz:2018ff}. 

As introduced above, researchers have extensively exploited the basin stability as an iconic measure to quantify the synchronization stability of power-grid nodes~\cite{Schultz:ch}. However, there is another crucial piece of information affecting the dynamical behaviors of phase oscillators; the basin stability, just as the order parameter of the conventional synchronization model such as the Kuramoto model~\cite{KuramotoModel}, depends on the transmission capacity between nodes (analogous to the coupling constant as a control parameter in the Kuramoto model, the details of which are presented in Sec.~\ref{sec:swing_equation}). For the past couple of years, we have actively participated in the process of incorporating the effects of transmission capacity into the framework of basin stability to aim at a comprehensive understanding of the characteristics of basin stability~\cite{Kim:2015kg,Kim:2016kd,Kim:2018do}. Throughout the series of work, we have discovered the relationship between the basin stability and the node's inconsistency in the community membership~\cite{Kim:2015kg,CommunityInconsistency}, the non-monotonic behavior of the basin stability as a function of transmission capacity in small building blocks of power grids~\cite{Kim:2016kd}, and its possible connection to chaotic dynamics~\cite{Kim:2018do}.

Based on our experience and insight gained throughout the previous works, in this study, we take a step further toward more lateral understanding in the whole picture of synchronization stability of power grids. We begin with the introduction to the theoretical background from Sec.~\ref{sec:methods}, including the integrated measure of the basin (in)stability for continuously varying values of transmission capacity.

\section{\label{sec:methods}Methods}

\subsection{Synchronization dynamics: the swing equation}
\label{sec:swing_equation}

The phase synchronization dynamics between power-grid nodes is usually described by the second-order Kuramoto-type model with inertia, so-called the swing equation~\cite{Filatrella:2008co,Dorfler:2013ew,Ji:2014ina,Nishikawa:2015gl}: 
\begin{equation}
\ddot \theta_i = \dot \omega_i = P_i - \alpha_i \dot \theta_i - \sum_{j} K_{ij} A_{ij} \sin(\theta_i - \theta_j) \,,
\label{eq:Kuramoto_type_equation}
\end{equation}
where the angular variable $\theta_i$ denotes the phase difference of node $i$ to the reference frame that is rotating with the rated frequency.
The net power input at node $i$ is represented as $P_i$ such that a net power producer (consumer) has $P_i > 0$ ($P_i < 0$) and a node with $P_i = 0$ does not inject or extract any electric power to or from the grid but serves as a structural branching point called a junction.
The damping coefficient $\alpha_i$ for the energy dissipation is proportional to the deviation of the angular frequency $\dot \theta_i=\omega_i$ of node $i$ from the reference frame, and the symmetric coupling term $K_{ij}=K_{ji}$ corresponds to the transmission capacity of the transmission line between nodes $i$ and $j$.
The symmetric adjacency matrix elements $A_{ij}$ describe the binary connection structure between nodes as $A_{ij} = 1$ if a transmission line exists between nodes $i$ and $j$, and $A_{ij} = 0$ otherwise. 
Note that in the synchronous state, the average frequency deviation over all of the nodes vanishes such that $\dot \theta_i = \omega_i = 0$. This swing equation is widely used to analytically or numerically study the most essential aspects of electrical power transmission in power grids.

\subsection{The basin stability}

In real operation of power grids, their synchronous state can be disturbed by external perturbations in terms of the phase and the angular frequency of each node as a phase oscillator. Depending to the size of the perturbations, the system can either recover its synchrony or fall into a limit cycle with its synchrony broken. One can quantify the resilience of the system's synchronous state against the perturbations to each node. A widely-used measure is the basin stability $B$, defined as the fraction of the state space composed of the phase and the angular frequency for each nodal oscillator, from which the system recovers its synchrony~\cite{Menck:2014fn}. In practice, one usually estimates the basin stability numerically by sampling the state space and simulating the swing equation. 

It is important to understand that basin stability is a synchronization stability measure assigned to each node, from the entire system's dynamics. In other words, it quantifies the entire system's synchronization stability in the case that the perturbations are applied to the node. For instance, $B_i=1$ implies that the system always keeps its synchronization regardless of any disturbance to node $i$; when $B_i = 0$, any amount of disturbance to node $i$ breaks the entire system's synchrony.

In this study, in order to numerically measure the basin stability of node $i$, we perturb each node by taking the phase and angular frequency values from the configuration space $\theta_i \in [-\pi,\pi)$ and $\omega_i \in [-100,100]$, uniformly at random as our previous studies ($500$ combinations of $\theta_i$ and $\omega_i$ for each $K$ value)~\cite{Kim:2015kg,Kim:2016kd,Kim:2018do}. 
For simplicity, we fix the dissipation coefficient $\alpha_i = \alpha = 0.1$ for all of the nodes~\cite{Menck:2013bk,Schultz2014detours,Menck:2014fn}. 
The numerical implementation is based on the fourth-order Runge-Kutta method~\cite{nr:07} with the convergence criteria $|\omega_i | < 5\times10^{-2}$ for every node $i$~\cite{Schultz2014detours}.

\subsection{Integrated basin instability}

\begin{figure*}[t]
\hfill\includegraphics[width=0.9\textwidth]{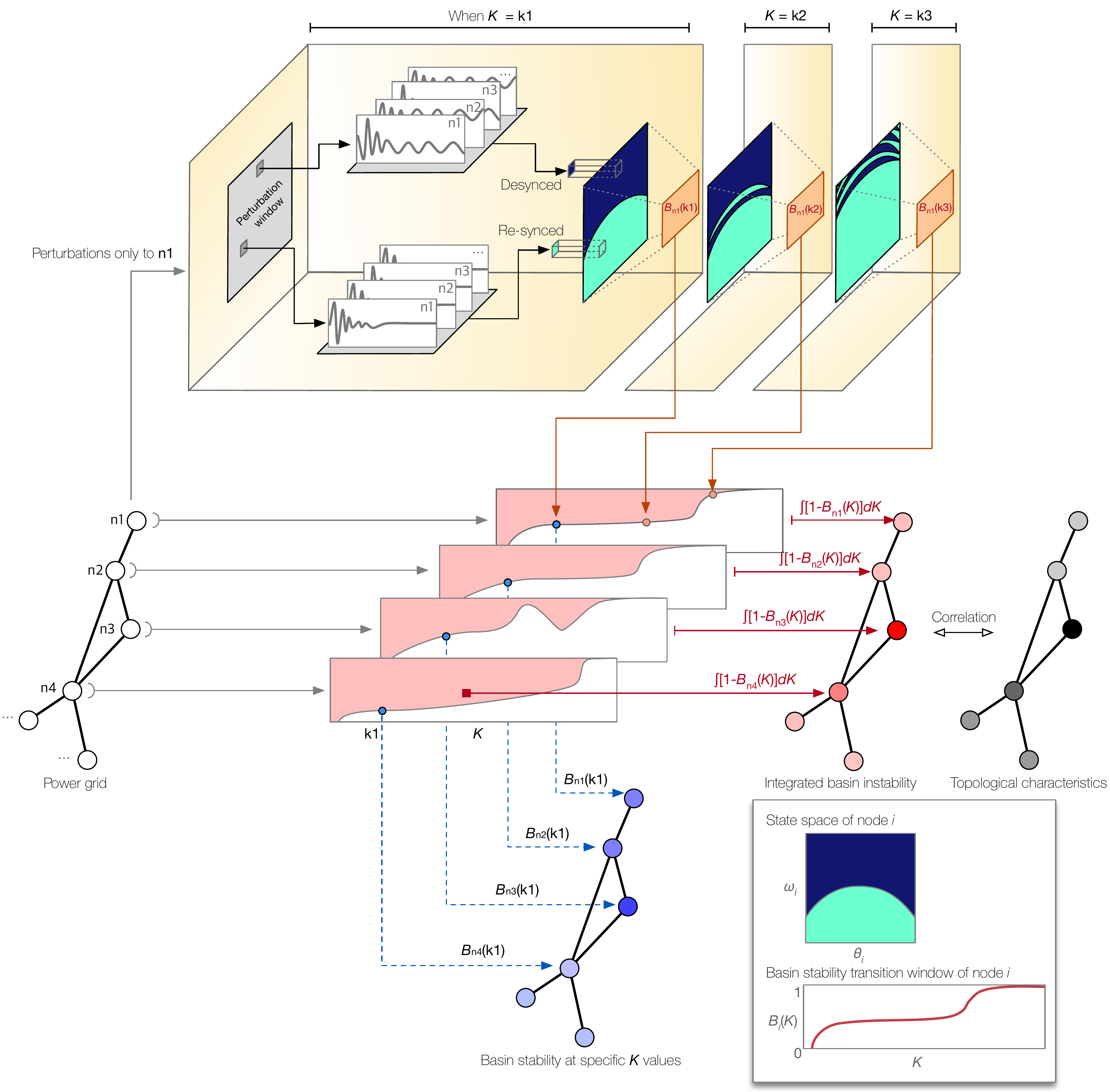}
\caption{In this illustration, we schematically show the overall procedure of measuring the basin stability and the integrated basin instability for each node.}
\label{fig:scheme}
\end{figure*}

Various parameters introduced in Sec.~\ref{sec:swing_equation}, such as the dissipation coefficient $\alpha_i$, the transmission capacity $K_{ij}$, and the net power input $P_i$, affect the basin stability. For instance, in general, the greater the dissipation coefficient is, the more easily synchronization occurs~\cite{Ji:2014ina}. In this work, we focus on the transmission capacity $K_{ij}$ of each connection between nodes $i$ and $j$, as the main control parameter. A power grid cannot maintain its synchronous state when the transmission capacity between its elements is too small, whereas once the transmission capacity exceeds a certain threshold, the system usually remains synchronized with the theoretically maximum basin stability of unity~\cite{Menck:2012vc,Menck:2014fn}. Such a transition pattern depending on the transmission capacity is also in accordance with the phase transition of the original Kuramoto model~\cite{KuramotoModel}. 

However, it was also reported that some systems show counterintuitive non-monotonic behaviors~\cite{Kim:2015kg,Kim:2016kd,Kim:2018do}. Therefore, we cannot simply find a single threshold value of transmission capacity by just detecting the ``jump'' of the basin stability and assume that the threshold value would characterize the system completely. Instead, we have to carefully scan a wide range of transmission capacity to obtain the whole picture of the transition pattern of the basin stability. Of course, it would be the best to show the whole transition shape without any information loss, as done in our previous works~\cite{Kim:2015kg,Kim:2016kd,Kim:2018do}, but it is cumbersome to take the entire functional shape as a measure of stability, e.g., when we try to compare it with other topological or dynamical characteristics of power grids. 

Motivated by such a trade-off between the accuracy in description and practicality, we introduce a collective measure of basin stability encompassing the entire range of transmission capacity. One could simply integrate $B_i$ from $K=0$ until the threshold value of $K$ at which $B_i(K)$ reaches unity.
However, the threshold $K$ value is distinct for each node, so setting the different upper bound for every node can be tricky and not practical.
Instead, we use the integral of $1-B$ (the pink shaded area over the curves in the central part of Fig.~\ref{fig:scheme}) as the \emph{integrated basin instability} (IBI):
\begin{equation}
C_i=\int_0^\infty \left[ 1-B_i(K) \right] dK \,,
\label{eq:ibs}
\end{equation}
where the convergence of $C_i$ is guaranteed as long as $B_i(K) = 1$ for $K$ larger than a certain finite threshold value.

Figure~\ref{fig:scheme} shows the schematic diagram of the entire process of measuring $C_i$, from the pedagogical description of the basin stability itself. Given a power grid, one can perturb a node by applying the initial perturbation uniformly at random from the state space composed of the phase and angular frequency for each $K$ value. The fraction of the configuration space that enables the recovery of synchronization is the basin stability of node $i$, denoted by $B_i(K)$ (corresponding to the area filled with the light cyan color in the top diagram of Fig.~\ref{fig:scheme}). 
We illustrate the process of measuring basin stability of node $\mathsf{n1}$ at $K=\mathsf{k1}$, $\mathsf{k2}$, and $\mathsf{k3}$, which is denoted by $B_{\mathsf{n1}}(\mathsf{k1})$, $B_{\mathsf{n1}}(\mathsf{k2})$, and $B_{\mathsf{n1}}(\mathsf{k3})$, respectively. 
Adjusting the $K$ values from $K=0$ to a large $K$ value exceeding the threshold, we obtain the transition curve of $B_i(K)$, which we call the transition window~\cite{Kim:2015kg}.
As the final step to calculating the IBI, we perform the integration of $1-B_i(K)$ over the range of $K$, or the area over the transition window. In this study, we scan $K$ by every $\Delta K=0.1$ and numerically perform the integration based on the composite trapezoidal method~\cite{Atkinson1989} without the error term:
$\int_a^b f(x)dx \approx \sum_{j=1} [f(x_{j-1})+f(x_j)] \Delta x /2$, where $\Delta x = 0.1$. By using the power-grid models that we will introduce from now on, we will present topological and dynamical factors to determine the IBI.

\begin{figure}[t]
\includegraphics[width=0.85\columnwidth]{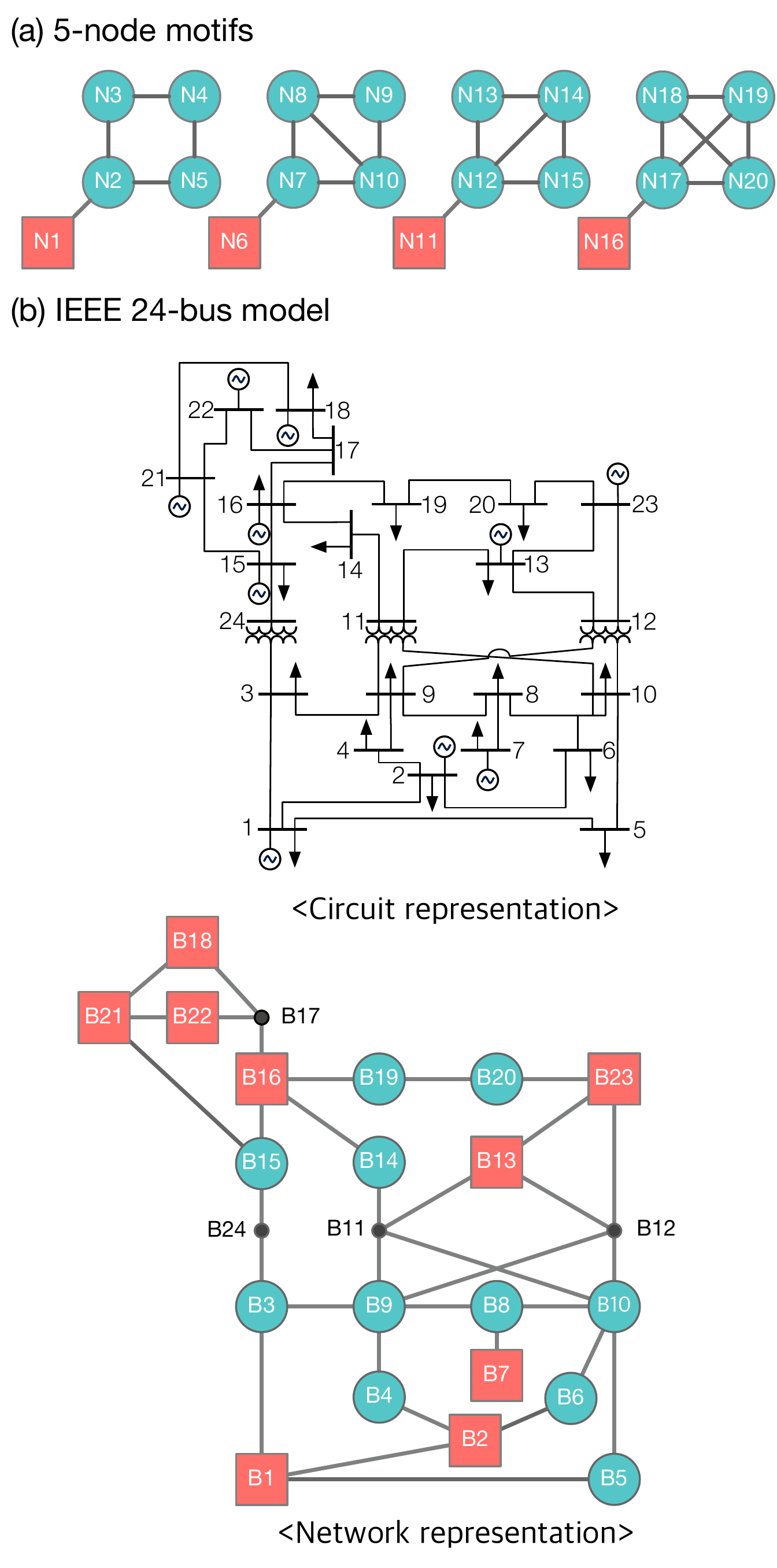}
\caption{The power-grid model networks we use in this work. (a) The 5-node motifs consist of a single power producers connected to a single consumer, with all of the possible connection configurations between the four consumers. The power producers (the coral red rectangles) supply four units of power to satisfy the a unit of demand of the four consumers (the turquoise blue circles). (b) The original IEEE 24-bus model above consists of $10$ power generators, $17$ loads, and $3$ transformers. We convert it to the network model below with the $9$ net producers (the coral red rectangles), the $11$ net consumers (the turquoise blue circles), and the $3$ junction nodes (the dark gray points). For both models, the codes in each node indicate the indices used in the main text.}
\label{fig:motifs}
\end{figure}

\section{\label{subsec:powergridmodels}Power-grid models}

A power grid is usually a single connected component consisting of power producers, consumers, and junctions.
However, the supply-demand relationship between power producers and consumers can be locally formed to establish small self-sustaining clusters in practice~\cite{Mureddu:2016dw}. Such small clusters can be operated autonomously even when other parts of the power grid are damaged, which in fact supports the resilience of the entire power grid in return. Therefore, small power-grid motifs can be taken as the minimal functional unit of power grids~\cite{Kim:2016kd,Kim:2018do}. In this spirit, we use two types of simple model networks for our comprehensive investigation of the factors affecting the synchronization stability: 5-node motifs as the basic building blocks of power grids~\cite{Kim:2016kd} and the IEEE 24-bus model as a real power grid, whose compactness enables the extensive analysis. 

First, we take the four distinct forms of 5-node motifs as shown in Fig.~\ref{fig:motifs}(a), which enumerate all of the possible cases of the four interconnected consumers and a single producer attached to one of the consumers. We choose this particular example to focus on the effect of the bottleneck or gate-keeper. 
For the balance of power production and consumption, the power producer has the net input power $P_i=4$ and the four consumers have the net output power $P_i=-1$, which satisfies Kirchhoff's circuit law of power conservation, i.e., $\sum_i P_i=0$. 
For the 5-node motifs, we measure the basin stability values for the transmission capacity uniformly assigned to all of the edges, in the range $0\leq K \leq 80$.

As a more realistic power grid model, we use the IEEE reliability test system (RTS)~\cite{Subcommittee:1979ih,Grigg:1999he}. The IEEE RTS is a conventional set of model power systems for testing various aspects of power grids in engineering, including the stability analysis of power grids~\cite{Arianos:2009ua,Bompard:2009ga,Dong:2010fx,Chen:2010fo,Salmeron:2004ju,Dorfler:2013ew}.
In this study, we particularly use the updated single area RTS-96 with $24$ buses~\cite{Ordoudis:2016ue} that reflects the recent features of modern power grids such as small scale wind farms.
This IEEE RTS 24-bus system (IEEE 24-bus) consists of $24$ buses with $10$ generators, $17$ loads, and $3$ transformers. Each bus, denoted by the indices in the circuit representation in Fig.~\ref{fig:motifs}(b), is composed of a localized set of possibly multiple facilities of generators that provide power, loads that consume power, and transformers that relay power without any loss (ideally).

In order to use the IEEE 24-bus data as a network representation composed of three types of nodes (producers, consumers, and junctions) that fits our framework, we treat individual buses as the nodes (we will call each bus a node accordingly), and assign the net input value $P_i$ to each node $i$ as the net sum of (positive) power generation and (negative) load belonging to the corresponding node. Each node in the network representation in Fig.~\ref{fig:motifs}(b) is indicated with the corresponding bus index (after `B') in the circuit representation. With this representation, we have $9$ net producers with $P_i > 0$ (the coral red rectangles in the bottom of Fig.~\ref{fig:motifs}(b)), $11$ net consumers with $P_i < 0$ (the turquoise blue circles), and $4$ junction nodes with $P_i = 0$ (the dark gray points), which constitute the $24$ nodes in total. 
We use the value of the maximum power in the original data~\cite{Ordoudis:2016ue} as the input power of the generators.
For the loads, since the original data~\cite{Ordoudis:2016ue} provides only the relative share of load distribution, we divide the total generation according to each of the relative fraction assigned to each load, and then set them as the amount of power consumption of the loads. Finally, as described above, we aggregate the total input power of the generators and the total power consumed at the loads belonging to node $i$, to set the net sum as the input parameter $P_i$ of node $i$. 
For the sake of computational cost, we rescale the net input values $P_i$ of the reference data by one tenth, while keeping the relative ratio between them.
See Appendix~\ref{sec:Appendix} for the actual $P_i$ values (dimensionless due to the rescaling) assigned to each node. 

In addition to the $P_i$ values taken from the real data as mentioned in the previous paragraph, we also make use of the real values of transmission capacity from the data source~\cite{Ordoudis:2016ue}. One crucial difference is that the transmission capacity is the control parameter we should adjust in the dynamical model. Therefore, we take the maximum transmission capacity values assigned to each edge described in the IEEE 24-bus data~\cite{Ordoudis:2016ue} and control the $K_{ij}$ values proportional to each of the maximum capacity value as the reference. 
More specifically, we use the scaling factor $r$ as the real control parameter multiplied to the reference transmission capacity $\kappa_{ij}$ for each edge between nodes $i$ and $j$, such that $K_{ij}=r \kappa_{ij}$ where $0\leq r \leq 20$.
Again, we use the rescaled (dimensionless) value of transmission capacity for the sake of computation ($1/100$ of the actual values~\cite{Ordoudis:2016ue} in the unit of megawatt in this case).
See Appendix~\ref{sec:Appendix} for the actual values of $\kappa_{ij}$ used in this study. 

With this setup, we obtain the power flow at each transmission line of the 5-node motifs and the IEEE 24-bus network based on the phase angle of each node at the synchronous steady state: 
\begin{equation}
F_{ij}=K_{ij}\sin(\theta_i-\theta_j) \,,
\label{eq:flow}
\end{equation}
where $F_{ij}$ is the current flow from $i$ to $j$. The power flow helps us to see the overall flow direction and is related to the current flow betweenness centrality (CFBC)~\cite{Newman:2005em}.
We cross-check the current flow from the swing equation with the power flow obtained from the Python package PyPSA~\cite{PyPSA}, and the results are the same. In practice, the PyPSA package takes the electrical reactance $X_{ij}$ as an input parameter instead of the transmission capacity $K_{ij}$. The two quantities are related as $K_{ij}= V_i V_j /  X_{ij}$ where $V_i$ is the amplitude of AC voltage on node $i$\cite{Lee:2017dz}. As the default setting of the amplitude is uniformly assigned as $V_i=1$ for all of the nodes, we utilize the relation $X_{ij} = 1/K_{ij}$ as the voltage information is missing in the data\footnote{The problem gets a bit trickier for the IEEE 24-bus network, as the data~\cite{Ordoudis:2016ue} itself includes the variable called ``reactance'' and it is \emph{not} inversely proportional to the transmission capacity variable in the same data. This may indirectly indicate the voltage information at each node, but we choose to stick to the relation $X_{ij} = 1/K_{ij}$ for the sake of simplicity.}.

\section{Results}
\label{sec:results}

\begin{table}[b]
\caption{\label{table:correlation}Three types of correlation coefficients and the $p$-value (corresponding to the null hypothesis of no correlation) between $C_i$ and dynamical and structure variables: $|P_i|$, degree, CFBC, and CM.} 
\begin{ruledtabular}
\begin{tabular}{c|c|r|l}
\multicolumn{1}{c|}{Type}  & $C_i$ vs & Coefficient & \multicolumn{1}{c}{the $p$-value}      \\\hline
\multirow{4}{*}{Pearson}   & $|P_i|$        & $0.837050$                & $3.416144\times 10^{-7}$ \\
                          & Degree         & $-0.052311$               & $8.081958\times 10^{-1}$ \\
                          & CFBC           & $-0.166695$               & $4.362666\times 10^{-1}$ \\
                          & CM             & $-0.123147$               & $5.664562\times 10^{-1}$ \\\hline
\multirow{4}{*}{Spearman} & $|P_i|$        & $0.866858$                & $4.282807\times 10^{-8}$ \\
                          & Degree         & $-0.102113$               & $6.349359\times 10^{-1}$ \\
                          & CFBC           & $-0.196564$               & $3.572674\times 10^{-1}$ \\
                          & CM             & $-0.103478$               & $6.303951\times 10^{-1}$ \\\hline
\multirow{4}{*}{Kendall}  & $|P_i|$        & $0.708315$                & $1.574424\times 10^{-6}$ \\
                          & Degree         & $-0.076613$               & $6.348642\times 10^{-1}$ \\
                          & CFBC           & $-0.134301$               & $3.585951\times 10^{-1}$ \\
                          & CM             & $-0.050725$               & $7.283948\times 10^{-1}$ \\
\end{tabular}
\end{ruledtabular}
\end{table}

\begin{figure}[t]
\includegraphics[width=0.85\columnwidth]{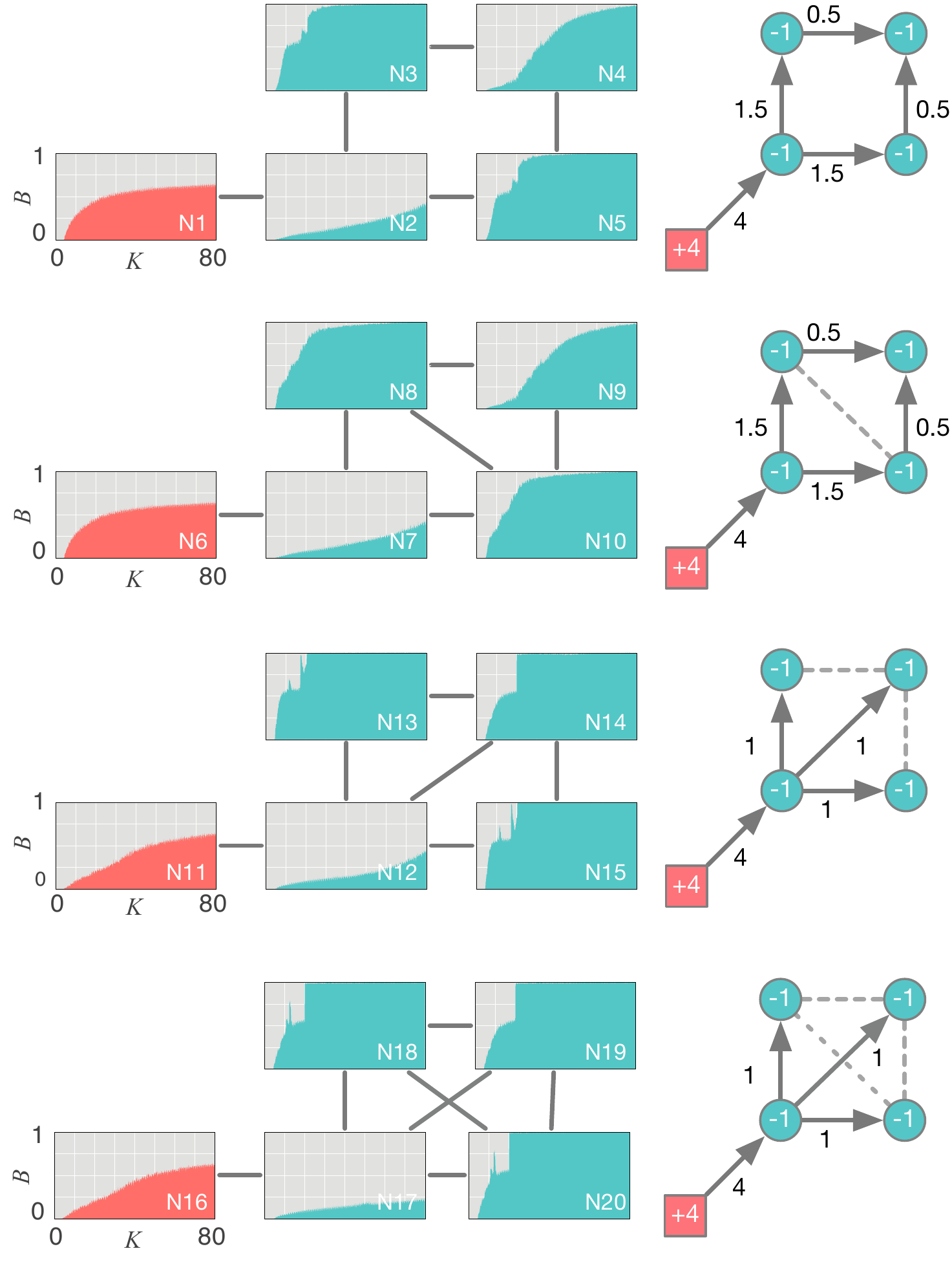}
\caption{The basin stability transition of the 5-node motifs and the current flow at the steady state. On the left side, the basin stability transition window inside each node, whose index corresponds to each node in Fig.~\ref{fig:motifs}(a), is presented for $0\leq K \leq 80$, where the network structure is illustrated between the nodes outside the transition windows as well. The direction and amount of positive current flow at the synchronous steady state is marked on the right side with the arrows and the numbers on the edges, while the undirected dashed lines represent the inactive links without any power flow. The flow direction is from $K=20$, but it is not altered within the range $6\leq K\leq80$. The coral red nodes (N1, N6, N11, and N16) are the power producers and the turquoise blue nodes (N2--N5, N7--N10, N12--N15, and N17--N20) are the consumers.}
\label{fig:motifsresult}
\end{figure}

From the the patterns of the basin stability transition of the 5-node motifs shown in Fig.~\ref{fig:motifsresult}, we find that both the topological structure and the power flow affect the transition pattern. On the left side, the transition window of each node is illustrated inside the node corresponding to the position in the topology, also marked by the node indices shared by Figs.~\ref{fig:motifs}(a) and \ref{fig:motifsresult}. In each motif, the producer node is located on the most left side (the coral red color) and the consumer nodes are on the right side (the turquoise blue color). For all of the four configurations, the producer nodes fail to reach unity until $K=80$ whereas most of the consumer nodes have reached $B=1$ at $K<80$ except for the nodes that connect the producer and the consumer-node group (N2, N7, N12, and N17). Considering the disproportionately large share of the power production concentrated to a single node (four times as much as the power consumption at each consumer node), we can conclude that the value of $P_i$ distributed throughout a power-grid network seems to affect the synchronization stability.

\begin{figure*}[t]
\includegraphics[width=0.9\textwidth]{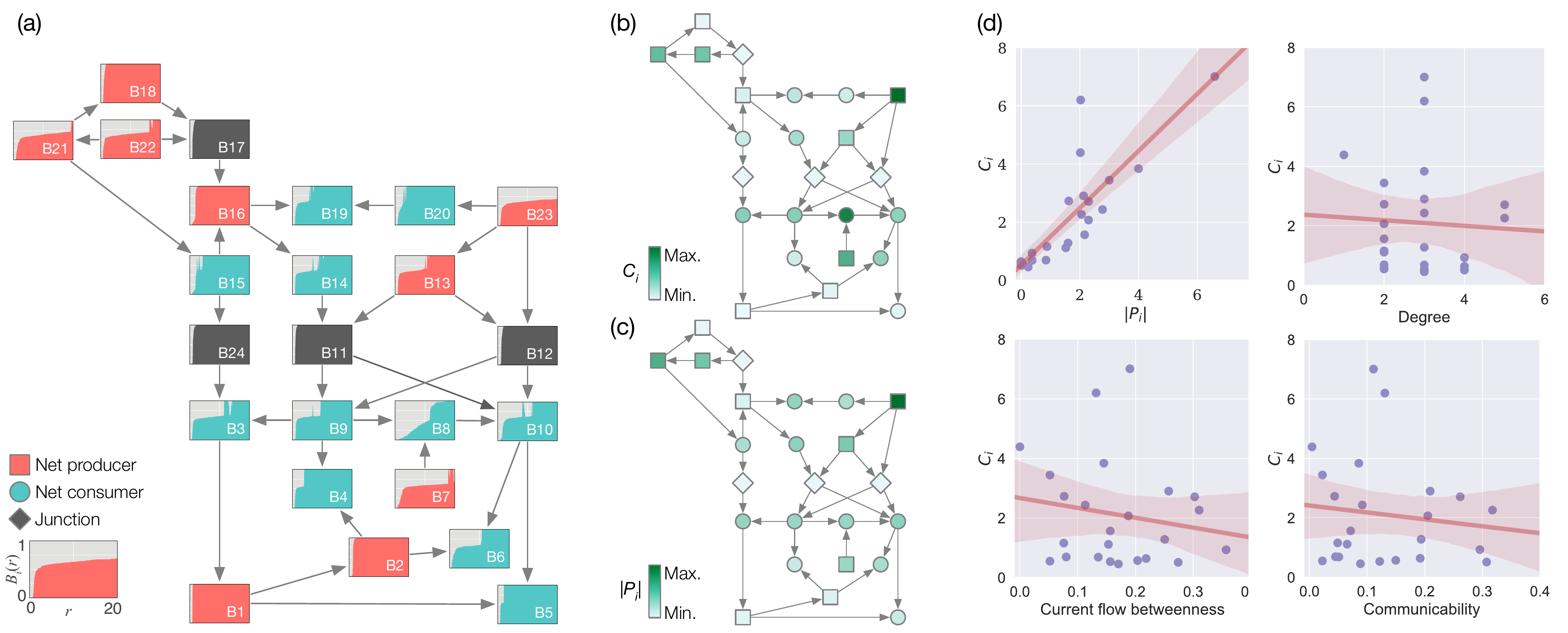}
\caption{The basin stability transition of the IEEE 24-bus network. (a) Each panel inside each node represents the basin stability transition window of the node for $0\leq r \leq 20$. All of the panels share the legend shown at the bottom left. The panels (b) and (c) show the network illustration of IEEE 24-bus with the color-coded values of $C_i$ and $|P_i|$, respectively. The arrow between the nodes in the panels (a), (b), and (c) indicates the direction of power flow. (d) The values of $C_i$ are correlated with the absolute value of input/output power $|P_i|$ more strongly with the Pearson correlation coefficient $0.84$ and the $p$-value $3.4\times10^{-7}$ (corresponding to the null hypothesis of no correlation) than any other network measures such as degree, current flow betweenness centrality (CFBC), and communicability (CM).}
\label{fig:IEEEresult}
\end{figure*}

On the right side of Fig.~\ref{fig:motifsresult}, we show the direction and amount of current flow at the steady state of each motif. The four units of power in total is distributed to the four consumer (sink) nodes to which one unit of power per node is consumed. Note that not all of the lines are used to transfer power, and the inactive edges without any power flow are represented by the undirected dashed lines.  In all of the four motifs, the power flows from the producer to the consumers through a single intermediary \emph{gate-keeper} nodes (N2, N7, N12, and N17). It is interesting to see that the basin stability of those gate keeper nodes are notably small throughout the whole range of $K$ values compared to the rest of the consumer nodes, which we call the gate-keeper effect. Since the gate-keeper nodes have to handle the lion's share of power flow, they play a crucial role in terms of the power distribution. Therefore, it is intuitively understandable that the perturbations on those nodes are likely to cause severe consequences.

We show that the two main results from the 5-node motifs: the effects of the input power $P_i$ and the gate-keeper location are also observed in the IEEE 24-bus model. Figure~\ref{fig:IEEEresult}(a) shows the transition window of the basin stability of each node in the IEEE 24-bus network. The nodes in the network show distinct transition patterns as the rescaled transmission capacity parameter $r$ varies, because each node has its own values of power input/output and a unique topological attribute. For instance, perturbation to some nodes, B1, B2, B11, B12, B17, and B24 shown in Fig.~\ref{fig:IEEEresult}(a), barely destroys the global synchronization even at relatively small $r$. On the other hand, B23 does not achieve $B=1$ even at the maximum transmission capacity $r=20$ in our simulation, at which most of nodes reach $B=1$. This unique behavior of B23 yields the largest IBI value $C_{i = \mathrm{B23}}$ defined in Eq.~\eqref{eq:ibs}, for that particular node. Note that the node B23 also has the largest input among all of the nodes as listed in Appendix. Besides this single case, we statistically find that the IBI values are correlated with the absolute value of power input/output values $|P_i|$ (with the Pearson correlation coefficient $0.84$ and the $p$-value $3.4\times10^{-7}$) more strongly than with any other network properties we examine such as degree, CFBC~\cite{Newman:2005em}, and communicability (CM)~\cite{Estrada:2008eo}, as shown in Figs.~\ref{fig:IEEEresult}(b) and ~\ref{fig:IEEEresult}(c), and Table~\ref{table:correlation}.

\begin{figure*}[t]
\includegraphics[width=0.9\textwidth]{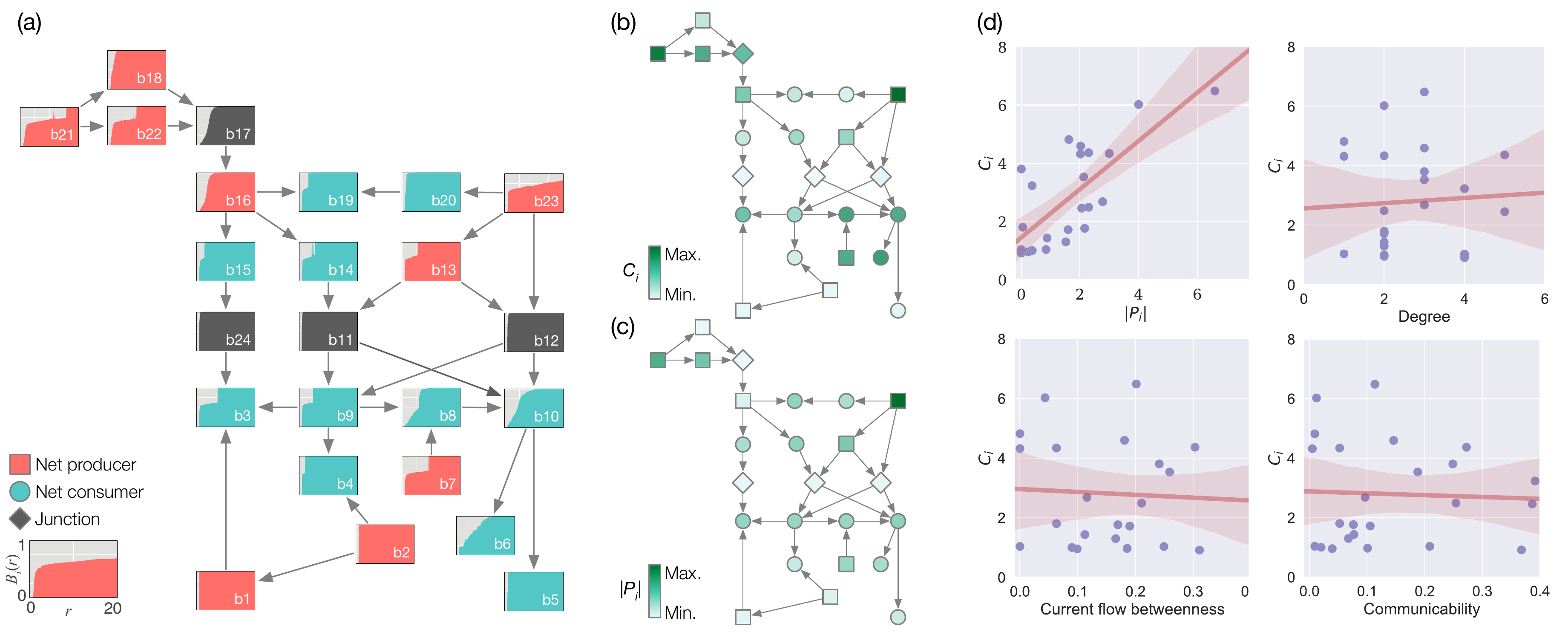}
\caption{The basin stability transition of the modified version of the original IEEE 24-bus network. All of the plots show the same type of information with Fig.~\ref{fig:IEEEresult}. The values of $C_i$ and $|P_i|$ show strong correlation with the Pearson correlation coefficient $0.747758$ and the $p$-value $2.7\times10^{-5}$ (corresponding to the null hypothesis of no correlation). The gate-keeper effect appears at b10 and b17, if we compared them with B10 and B17 in Fig.~\ref{fig:IEEEresult}.}
\label{fig:IEEEmodifiedresult}
\end{figure*}

In contrast to the 5-node motifs, the original IEEE 24-bus model does not have such an extreme case of gate-keeper nodes. To validate the gate-keeper effect observed in the 5-node motifs, we slightly modify the IEEE 24-bus model by eliminating three edges to enforce the gate-keeper structure: the edges between B1 and B5, between B2 and B6, and between B15 and B21 (see Fig.~\ref{fig:IEEEmodifiedresult} for the modified structure). As a result of the removal of the three edges, we indeed find the gate-keeper effect. If we compare the nodes b10 and b17 to their counterparts (B10 and B17) in the original network in Fig.~\ref{fig:IEEEresult}, their IBI values of the former are notably larger than the latter as a result. In particular, the anomalous peak~\cite{Kim:2018do} in B10 disappears in b10, which indicates the overall change in the transition pattern. The strong correlation between $C_i$ and $|P_i|$ still remains, as shown in Fig.~\ref{fig:IEEEmodifiedresult}(d) with the Pearson correlation coefficient $0.747758$ and the $p$-value $2.7\times10^{-5}$.

\section{Summary and conclusions}
\label{sec:conclusion}

We have investigated the transition pattern of basin stability to understand the landscape of synchronization stability of power-grid nodes in a comprehensive manner. By integrating the basin instability ($1-B$) to define the concept of the integrated basin instability (IBI), we have represented each node's instability across different values of transmission capacity. With the 5-node motifs and the IEEE 24-bus model as the representative cases of building blocks of localized power grids and real power grids in engineering, respectively, we find that the nodes with the larger values of input/output power tend to have the larger IBI, i.e., more unstable in terms of perturbation. It implies that concentration of disproportionately large amount of power generation at a single location in a power grid can be dangerous, as such a facility would be more fragile in response to external perturbation. This may sound intuitive, but we would like to emphasize that such heterogeneous distribution of input/output power besides the topological characteristics has not been systematically investigated under the condition of varying transmission capacity in many previous studies.
 
On the topological side as well, the dynamical variable of the current flow is deeply related to the synchronization stability. For instance, we have found that perturbation to nodes located at the position where the current flow is concentrated easily break the system's synchronization. This gives a hint to improve the power grid's stability, e.g., designers of power grids should prepare bypasses for electric current to flow in various paths to avoid a small number of gate-keeper nodes.

As the first step toward the understanding of more real aspects of power grids, our result is based on one of the power-grid models of the IEEE reliability test system. More rigorous model analyses both in terms of quality and quantity would help to understand the general relationship between the topology and synchronization stability of power grids as it has been done in the DC power grid~\cite{Wienand:2019wk}. Another challenge is the fact that the calculation of IBI requires extensive numerical costs, so the development of more efficient algorithms to estimate the comprehensive basin stability is necessary, which can also be one of the future directions of this important research topic affecting our entire civilization.

\begin{acknowledgments}
This work was supported by the National Research Foundation of Korea (NRF) Grant No. NRF-2018R1C1B5083863 (SHL) and NRF-2017R1D1A1B03032864 (S-WS). 
\end{acknowledgments}

\begin{appendix}
\section{Actual values of the parameter used in this study}
\label{sec:Appendix}

In this Appendix, we show the detailed values of the input power of the nodes and the transmission capacity of the links that we used in this study in Tables~\ref{table:IEEE_node} and \ref{table:IEEE_link}.

\begin{table}[t]
\caption{\label{table:IEEE_node}The input power $P_i$ in an arbitrary unit and the role of each node in the IEEE 24-bus network. The values are taken from the original data~\cite{Ordoudis:2016ue}. The node indices correspond to the ones (after `B') in the network representation shown in Fig.~\ref{fig:motifs}.}
\begin{ruledtabular}
\begin{tabular}{c|c|c||c|c|c}
Node & $P_i$        & Role     & Node & $P_i$       & Role     \\ \hline
1     & 0.2375   & Producer & 13    & 2.77125  & Producer \\
2     & 0.3725   & Producer & 14    & $-2.295$   & Consumer \\
3     & $-2.12625$ & Consumer & 15    & $-1.59625$ & Consumer \\
4     & $-0.8775$  & Consumer & 16    & 0.36875  & Producer \\
5     & $-0.84375$ & Consumer & 17    & 0      & Junction \\
6     & $-1.62$    & Consumer & 18    & 0.05125  & Producer \\
7     & 2.015    & Producer & 19    & $-2.16$    & Consumer \\
8     & $-2.025$   & Consumer & 20    & $-1.51875$ & Consumer \\
9     & $-2.05875$ & Consumer & 21    & 4.0      & Producer \\
10    & $-2.295$   & Consumer & 22    & 3.0      & Producer \\
11    & 0      & Junction & 23    & 6.60     & Producer \\
12    & 0      & Junction & 24    & 0      & Junction \\ 
\end{tabular}
\end{ruledtabular}
\end{table}

\begin{table}[t]
\caption{\label{table:IEEE_link}The reference transmission capacity $\kappa_{ij}$ in an arbitrary unit assigned to each edge between $i$ and $j$. The values are taken from the original data~\cite{Ordoudis:2016ue}. The node indices correspond to the ones (after `B') in the network representation shown in Fig.~\ref{fig:motifs}.}
\begin{ruledtabular}
\begin{tabular}{cc|c||cc|c}
$i$ & $j$        & Capacity    & $i$ & $j$        & Capacity     \\ \hline
1     & 2   & 1.75 & 11    & 13  & 5.00 \\
1     & 3   & 1.75 & 11    & 14   & 5.00 \\
1     & 5 & 3.50 & 12    & 13 & 5.00 \\
2     & 4  & 1.75 & 12    & 23  & 5.00 \\
2     & 6 & 1.75 & 13    & 23      & 5.00 \\
3     & 9    & 1.75 & 14    & 16  & 5.00 \\
3     & 24    & 4.00 & 15    & 16    & 5.00 \\
4     & 9   & 1.75 & 15    & 21 & 10.00 \\
5     & 10 & 3.50 & 15    & 24      & 5.00 \\
6    & 10   & 1.75 & 16    & 17      & 5.00 \\
7    & 8      & 3.50 & 16    & 19     & 5.00 \\
8    & 9      & 1.75 & 17    & 18      & 5.00 \\
8    & 10      & 1.75 & 17    & 22      & 5.00 \\
9    & 11      & 4.00 & 18    & 21      & 10.00 \\
9    & 12      & 4.00 & 19    & 20      & 10.00 \\
10    & 11      & 4.00 & 20    & 23      & 10.00 \\
10    & 12      & 4.00 & 21    & 22      & 5.00 \\
\end{tabular}
\end{ruledtabular}
\end{table}
\end{appendix}

\bibliography{main_revised}

\end{document}